# On the semi-annual, 27 day, variation in geomagnetic activity, cloud cover and surface temperature.

I. R. Edmonds
12 Lentara St, Kenmore, Brisbane, 4069 Australia.
ian@solartran.com.au

**Abstract.**
We develop a basic model of the time variation of geomagnetic activity and show that the model predicts, with decreasing levels of exactitude, the time variation of the ~27 day period components of geomagnetic aa index, cloud cover and surface temperature during several years near solar cycle minima. We interpret this as indicating that there is a connection between the ~27 day variation of geomagnetic activity and the ~27 day variations of cloud cover and surface temperature with the decreasing correlations between model variation and aa index, cloud cover and surface temperature variation due to delays and phase shifts between the three variables some of which are obvious, such as the 180$^{o}$ seasonal phase shift between cloud cover and surface temperature, and others less certain. We find that, while the components of cloud cover and surface temperature influenced by geomagnetic activity amount to, on average, about 20% of the overall variations, the influence may be several times higher during the semi-annual maxima in geomagnetic activity that occur around the equinoxes.

## 1. **Introduction.**

The semi-annual effect in geomagnetic indices is generally attributed to one or more of three mechanisms: the axial or Rosenberg-Coleman mechanism, the equinoctial mechanism, and the Russell-McPherron mechanism, Cortie (1912), Rosenberg and Coleman (1969), Russell and McPherron (1973), Cliver et al (2002), Cliver et al (2004). The mechanisms are associated with (a), the deviation of the Earth's orbit to higher solar latitudes around the equinoxes, (b), the angle between the Sun-Earth line and Earth's magnetic dipole or (c), some combination of (a) and (b), Cliver et al (2004). The difference in timing of the semi-annual maximum in the three different mechanisms is only about one month (7 March – 7 April) making it challenging to distinguish between the mechanisms, Cliver et al (2004).

The objective of this paper is to utilise a tilted solar dipole model to derive the time variation of geomagnetic activity and use this time variation as the basis for establishing that, in solar minimum years, there is a linear relationship between the ~27 day period variations of geomagnetic activity, cloud cover and surface temperature. There is an extensive literature on the origin of geomagnetic activity, e.g. Svalgaard (1977) and Svalgaard and Cliver (2007) and references therein, and on observations and theories relating to the connection between solar activity and climate variables e.g. reviews by Gray et al (2010), Kirby (2007), Tinsley et al (2007), Lockwood (2012), Rycroft et al (2008).

The arrangement of the paper is as follows: Section 2 outlines the data analysis method. In section 3 we derive a model of the semi-annual and 27 day variation of geomagnetic activity. Section 4 outlines the criteria for selecting specific years for the study. Section 5 demonstrates correlation between the model variation, geomagnetic aa index variation and variation of cloud cover at a mid-latitude southern hemisphere location. Section 6 demonstrates the correlation and phase relationship between aa index, cloud cover and surface temperature at the same location. Section 7 extends the study to northern hemisphere mid-latitude locations. Section 8 discusses the results, possibilities for future work and concludes that there is a significant semi-annual, 27 day variation in mid-latitude cloud cover and a significant semi-annual, 27 day variation in mid-latitude surface temperature connected to solar rotation.

## 2. Data analysis.

The data used in this study are records of geomagnetic aa index (aa), cloud cover, (CC) and surface temperature, (T). Isolation of the ~ 27 day component of each variable was obtained by making a Fast Fourier Transform of each data series. The resulting n Fourier amplitude and phase pairs, $A_n(f_n)$, $\phi_n(f_n)$, in the frequency range 0.031 days$^{-1}$ to 0.045 days$^{-1}$ (period range 32 to 22 days) were then used to synthesize a band pass filtered version of each variable, denoted for example 27T, by summing the n terms, $T_n = A_n\mathrm{Cos}(2\pi f_n t - \phi_n)$ for each day in the series. Where a data series has been smoothed by, for example, a 30 day running average, the resulting smoothed series is denoted e.g. T S30.

## 3. Model of the time variation of geomagnetic activity.

The introduction mentioned several mechanisms that predict the variation of geomagnetic activity with time. Here we outline a model of geomagnetic activity based on a two sector magnetic field that includes a semi-annual phase reversal of the field.

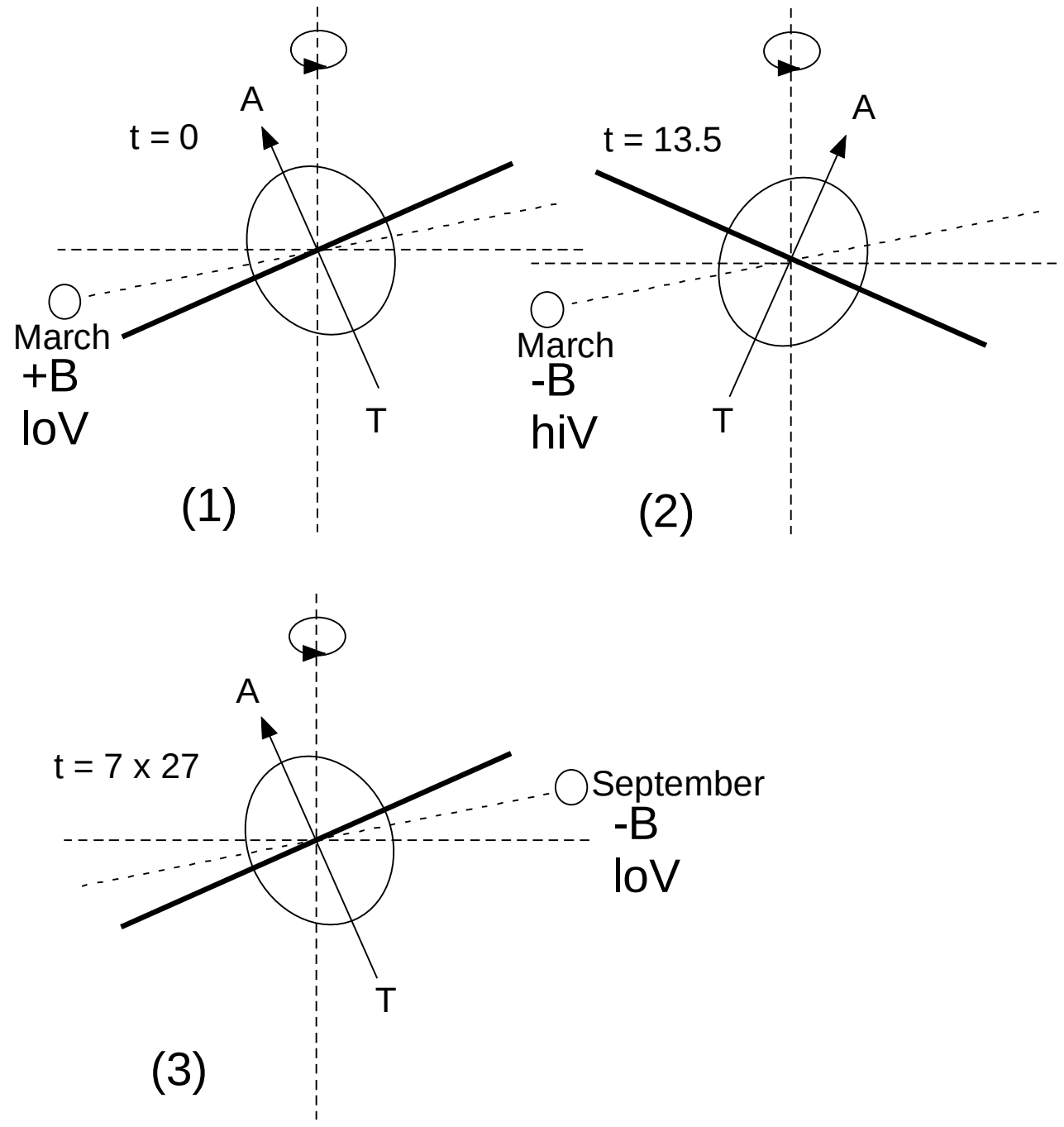

**Figure 1.** Sun – Earth geometry of the tilted solar dipole, flat heliospheric sheet, two sector model used to estimate the time variation of geomagnetic activity.

Figure 1 illustrates the Sun – Earth geometry associated with the semi-annual effect. The large circle represents the Sun, the vertical axis is the Sun's rotation axis, the arrow labelled A T is the direction of the Suns magnetic dipole here shown tilted from the rotation axis by about $30^o$, the thick line represents a flat heliospheric current sheet (HCS). The field above the HCS is radial and away from the Sun (A) and below the HCS is radial and toward the Sun (T). The Earth is represented by the small circle. The Earth rotates around the vertical axis with a period of 365 days. The synodic period of rotation of the Sun is taken as 27 days. Sketch (1) in Figure 1 shows the geometry at t = 0 when the Earth is at its highest southern heliographic latitude at March equinox, the Suns magnetic dipole is angled towards Earth and the Earth is in the away (A) magnetic field represented as +B and the solar wind velocity is low. Sketch (2) is 13.5 days later when the Earth is in the towards (T) magnetic field represented as –B and the solar wind velocity is high. Thus the magnetic field and the solar wind velocity at Earth vary with a period of 27 days. The axial and Rosenberg-Coleman models of the effect also represent the geometries in sketches (1) and (2) as giving rise to a strong 27 day period variation of the magnetic field and the solar wind velocity. Sketch (3) represents the geometry six months later in the year when the Earth is at its highest northern solar latitude at September equinox. At this time the amplitude of the 27 day variation of magnetic field and solar wind velocity is at the second semi-annual maximum. The times midway between March and September, i.e. times around the solstices, when the Earth is near zero solar latitude are the times of minimum amplitude of the 27 day variations in magnetic field and solar wind velocity. The explanations of the semi-annual effect according to the equinoctial and Russel-McPherron mechanisms differ from the above but give essentially the same semi-annual time variation, Cliver et al (2002).

One aspect of the semi-annual effect that appears to be neglected in most of the earlier descriptions is the phase change in the 27 day variation of the magnetic field that occurs when the Earth passes through zero heliographic latitude. Note that, relative to the 27 day period synodic timing, the sign of the magnetic field as depicted in sketch (3) at t = 7 x 27 days is the inverse of the sign of the magnetic field in sketch (1) at t = 0. Thus the magnetic field variation in the second (September) half of the year is phase shifted by $180^o$ relative to the phase of the magnetic field variation in the first (March) half of the year. However, note that the phase of the solar wind velocity variation is unchanged. The relative daily variation of the solar magnetic field (B) at Earth can therefore be represented by an amplitude modulation of the form $B = (+/-)\cos(2\pi(t - 91)/365)\cos(2\pi t/27)$ and the relative daily variation of solar wind velocity (V) by $V = 1 + 0.25\cos(2\pi t/27)$. Numerous studies e.g. Svalgaard (1977), Cliver and Svalgaard (2007), indicate that geomagnetic activity varies as $BV^2$ so we model the relative variation of geomagnetic activity as

$$\text{model} = BV^2 = (+/-)\cos(2\pi(t - 91)/365)\cos(2\pi t/27)[1 + 0.25\cos(2\pi t/27)]^2 \qquad (1)$$

where the +/- sign takes account of solar dipole reversal. Figure 2A shows the model variation over one year as well as the variation of the three components of the model. The components, obtained by band pass filtering as described in Section 2, are an annual component, a ~27 day component and a ~13.5 day component. Figure 2B shows the frequency spectrum of the model variation. We note that the annual component is $-0.25\cos(2\pi(t-91)/365)$ and the ~27 day component is the same as the B variation, (+/-

)cos(2π(t-91)/365)cos(2πt/27). In this paper we are primarily interested in ~27 day components and will use the ~27 day component of the model of geomagnetic activity in a slightly more general form:

$$27model = (+/-)\cos(2\pi(t-t_A)/365)\cos(2\pi(t-t_S)/27) \quad (2)$$

where 27model denotes the ~27 day component of relative geomagnetic activity, $t_S$ is the day of first solar rotation maximum and $t_A$ is the day of first semi-annual maximum. Figure 2A shows the model variation with $t_S$ = 0 and $t_A$ = 91 days. We note that the model variation is asymmetric between the first and the second equinox due to the fact that the magnetic field component experiences a 180$^o$ phase shift.

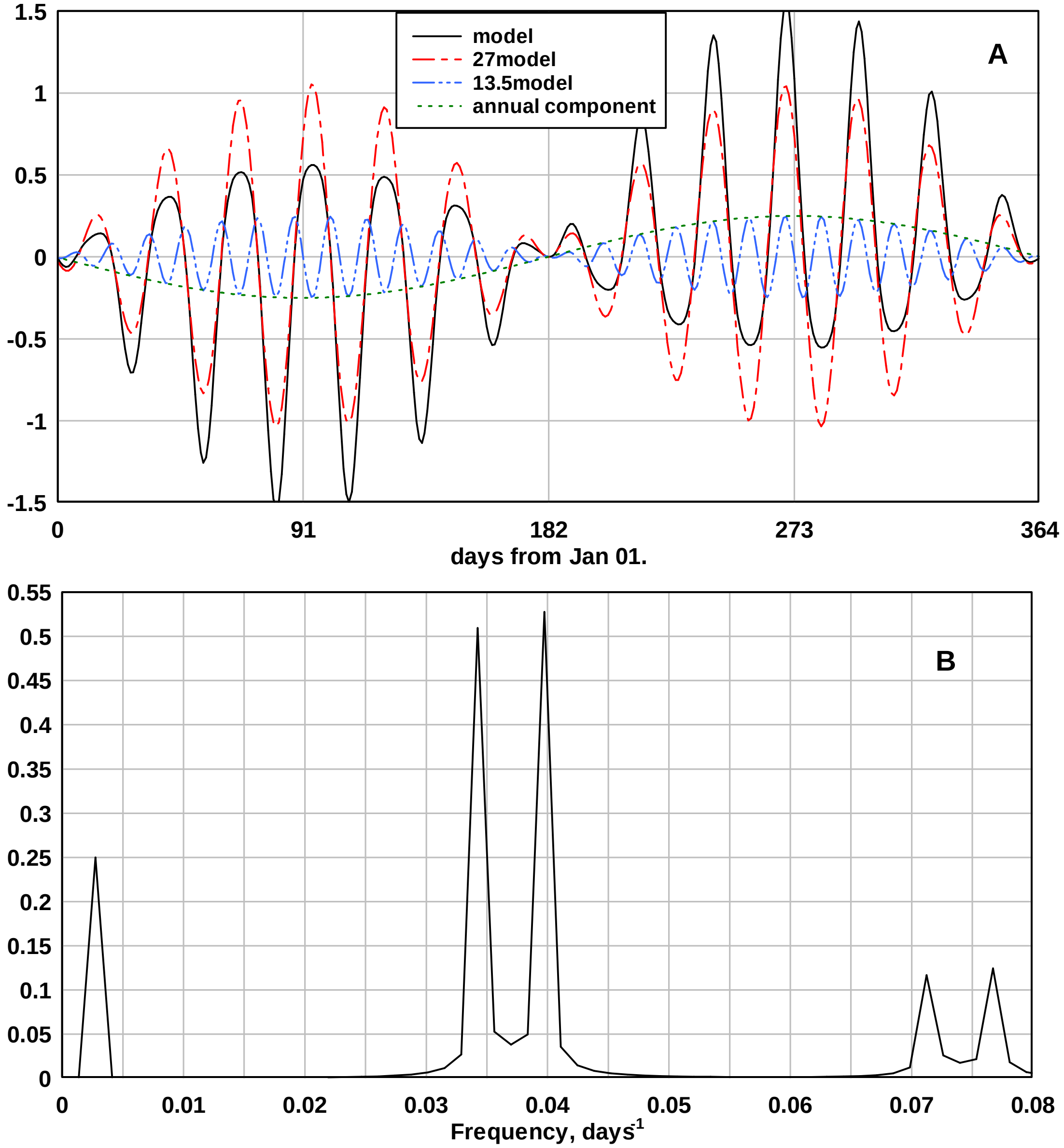

**Figure 2.** (A) variation of the model of geomagnetic activity and the annual, 27 day and 13.5 day components of the model variation. (B) The frequency spectrum of the model variation.

Equation (2) can be expanded to

$$27model = (+/-)[\cos(2\pi(f_S + f_A)t - (\phi_S + \phi_A)) + \cos(2\pi(f_S - f_A)t + (\phi_S - \phi_A))]/2 \qquad (3)$$

where $f_A$ = 1/365 days$^{-1}$, $f_S$ = 1/27 days$^{-1}$, $\phi_S = 2\pi f_S t_S$ radians and $\phi_A = 2\pi f_A t_A$ radians. Equation (3) is similar to the expression used by Coleman and Smith (1966) to interpret peaks in the frequency spectra of the Ci and Kp magnetic indices. The interpretation of the occurrence of minima in the Rosenberg-Coleman description is that, when Earth is at zero heliographic latitude, the probability of positive or negative magnetic fields occurring is equal so the average amplitude of observed magnetic field approaches zero. The semi-annual variation of the ~27 day component of the Ci index was observed by Shapiro (1969). Equations (2) and (3) can also be related to the phase modulation relation used by Takalo and Mursula (2002) to interpret peaks in the spectra of the Bx component of the interplanetary magnetic field. We note that, from equation (3), the frequency spectrum will contain two peaks, at frequencies $f_S$ +/- $f_A$ (periods 25 and 29 days). In the work that follows we will select for study years with semi-annual variations that have minima close to the first day of the year, day 1. Therefore we will usually set $t_A$ in equation (1) to 91 days placing the maxima at day 91 and day 273. However, $t_A$ may, occasionally, be varied by a few days from 91 days to provide a better fit of the model to the data.

The focus of this work will be on demonstrating that, for years near solar cycle minimum, the 27 day components of geomagnetic activity, cloud cover and surface temperature each vary in a similar manner to the model variation in Figure 2A and have a spectrum similar to that in Figure 2B.

**4. Selecting years of semi-annual variation to study.**

The semi-annual variation in geomagnetic activity is highly variable from year to year. Cliver et al (2004) found that only two years, 1954 and 1996, met their criteria for an "ideal" semi-annual variation in 130 years of aa geomagnetic index data. The daily cloud data for Hobart used in this study extends over 92 years from 1918 to 2010 and it is therefore necessary to select several years out of the 92 for study. The variable we use for selection is the daily maximum surface temperature anomaly at Hobart, available at http://www.bom.gov.au/climate/change/hqsites/data/temp/maxT.094029.daily.txt and is referred to here as the THOB anomaly. Figure 3 shows the 90 day moving average (S90 – S365) of THOB anomaly for the 20 year, 14600 day, interval beginning on January 01, 1918. The temperature anomaly results after the annual seasonal variation of temperature has been removed by the Bureau of Meteorology so the residual annual variation evident in Figure 3 must be due an effect different from the seasonal variation. We infer that the residual annual variation shown in Figure 3 is associated with the annual variation in geomagnetic activity shown in Figure 2A and this inference will be discussed in detail later. However, for the time being, this annual variation is a convenient way of selecting "ideal" years for study. In the present case "ideal" is taken to mean a symmetrical temperature anomaly variation that passes through zero on day one of the year i.e. similar in variation to the annual variation in Figure 2A. Years that approximate the "ideal" are indicated by letters a to h in Figure 3. Also indicated in Figure 3 is the 90 day moving average of the daily sunspot area. The daily average sunspot area 1874 to 2013 is available at http://solarscience.msfc.nasa.gov/greenwch/daily_area.txt
.Comparison of smoothed temperature anomaly and smoothed sunspot area makes it evident that "ideal" temperature anomaly variations occur predominantly during solar cycle minima. During solar cycle maxima the "ideal" temperature anomaly variation is disturbed by events e.g. x, y and z in Figure 3, associated with peaks in sunspot activity. Study of events during solar maxima is outside the scope of the present paper. Based on the criteria above we select years 1922-1923 (c in Figure 3), 1932-1933 (g in Figure 3), 1994 -1995 (a in Figure 3B supplement), and 2007-2008 (a in Figure 3C supplement) to study.

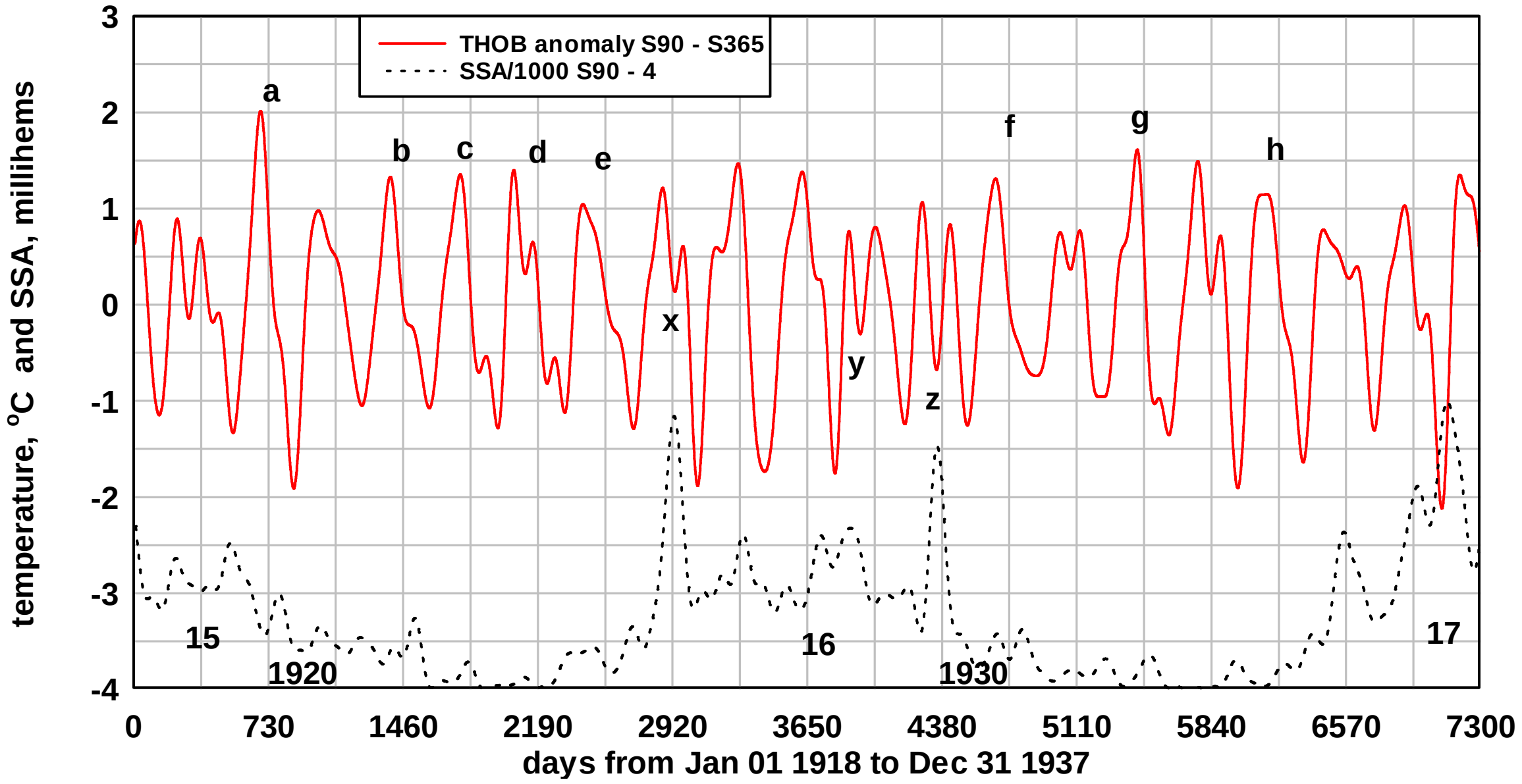

Figure 3. Shows the 90 day running average of the daily maximum temperature anomaly (THOB) at Hobart for the years 1918 – 1937. Also shown is 90 day running average of the sunspot area for solar cycles 15, 16 and 17. Lower case letters a – h mark symmetrical THOB anomaly variations that cross zero near day one of the year. Letters x – z mark times when the THOB anomaly is disturbed by peaks of sunspot activity.

## 5. The semi-annual variation in geomagnetic activity and cloud cover.

The daily average geomagnetic aa index 1862 - 2013 is available at ftp://ftp.ngdc.noaa.gov/STP/SOLAR_DATA/RELATED_INDICES/AA_INDEX/aaindex
The daily average cloud cover data for Hobart (station 094029) 1893 to 2010 was purchased from the Australian Bureau of Meteorology. Cloud data for a few years during World War 1 are missing so this study uses the data beginning on Jan 01, 1918. As we are primarily interested in the ~27 day variations, the data series for aa index and the cloud cover were band pass filtered as described in Section 2 to retain only components in the frequency range 0.031 to 0.045 days$^{-1}$ (32 to 22 day period). The model variation, equation (2), is compared with the ~27 day aa index variation during 1922 1923, (reduced by a factor of 10), in Figure 4A and with the ~ 27 day cloud cover variation during 1922 1923 in Figure 4B. The model was fitted to the data by eye by adjusting the solar rotation period slightly from 27 days and/or by adjusting $t_A$ slightly from 91 days or $t_S$ slightly from 0 days. In Figure 4 the model variation is 27model = (-1)cos(2π(t – 91)/365)cos(2π(t - 4)/27) and the parameters used in the model are indicated as (-1), 27, 91, 4 in the Figure 4 legend. A good correlation between the model variation and the variation of the ~27 day component of aa index is evident. The correlation coefficient over the entire two year interval is r = 0.54. However, the correlation coefficient in the interval of the two half years on each side of Jan 01 1923 is r = 0.63.

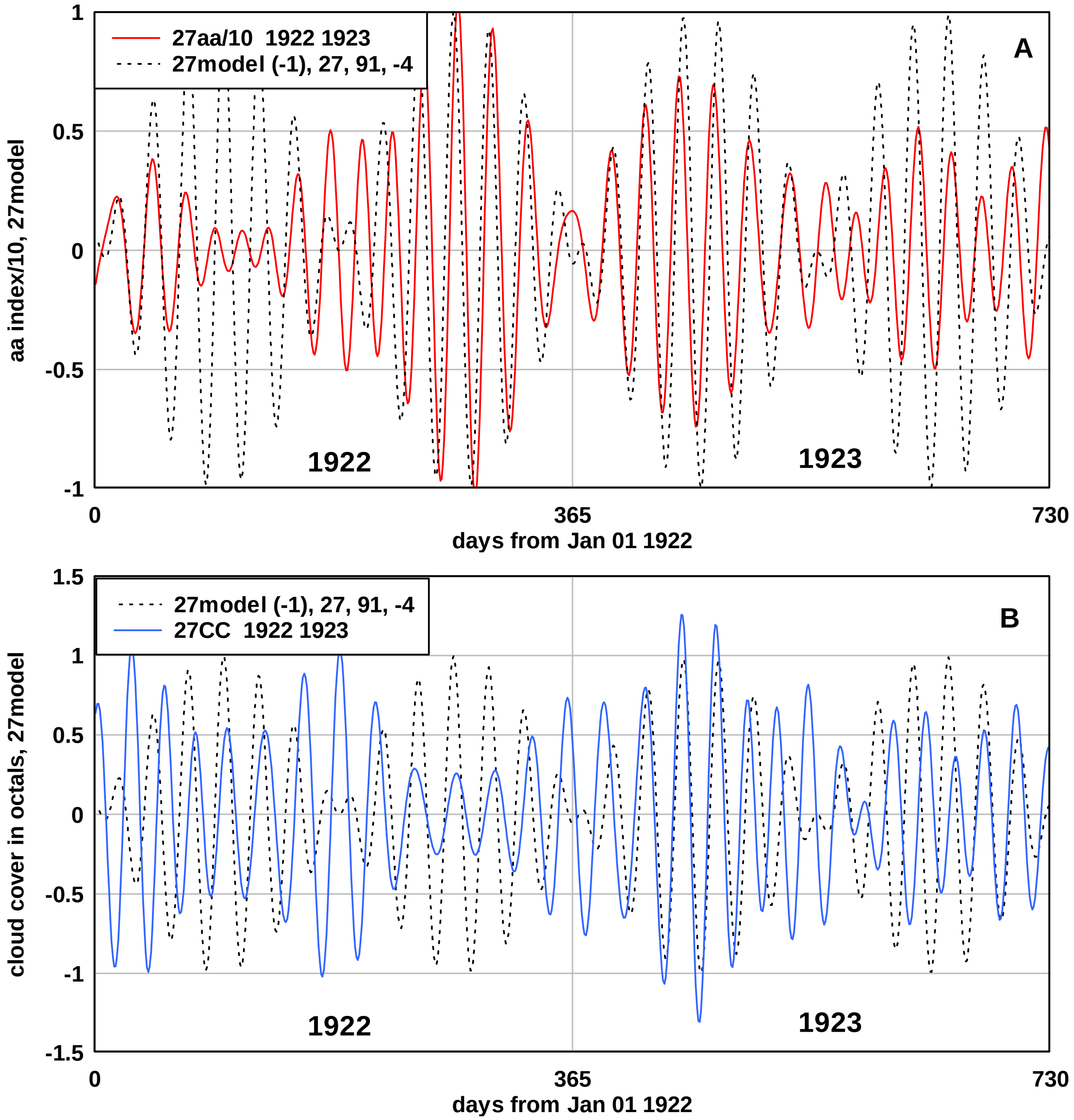

**Figure 4.** (A), compares the model variation with the variation of the ~27 day component of aa index (reduced by a factor of 10) for the years 1922 1923. (B), compares the model variation with the variation of the ~27 day component of cloud cover at Hobart for the years 1922 1923.

Figure 4B compares the variation of the ~27 day component of daily cloud cover at Hobart during 1923 1924 with the model variation. The correlation over the entire two year period is r = 0.27. However, the correlation over the interval of the two half years on each side of Jan 01 1923 is r = 0.49. Clearly the correlation with the model has decreased. We expect the correlation to decrease as we progress down the chain of connection i.e. from the model variation as depicted in Figure 2A, to the variation of geomagnetic activity, to the variation in cloud cover and, ultimately to the variation in surface temperature. Each step in the sequence is expected to be affected by inputs from sunspot activity, measurement noise (e.g. cloud cover is subjectively assessed in eights of the sky covered), synoptic weather noise, and time delays and phase shifts associated with the development of cloud cover and surface temperature.

Figure 5 compares the ~27 day variations of the aa index and the cloud cover at Hobart during 1932 1933 with the model. Here the semi-annual variation in amplitude of each of the variables is more clearly evident and the correlation between variables is higher. The measured correlation coefficients, r, over the entire two year interval are: 27model – 27aa index, r = 0.77; 27aa index – 27CC, r = 0.45; and 27model – 27CC, r = 0.29. It is noticeable that the cloud cover varied in-phase with the aa index during this two year interval apart from July, August and September in 1933. Similar comparisons of model, aa index and cloud cover for the intervals 1994-1995 and 2007-2008 are available in Figure 5B and 5C in the supplement. If the variation in geomagnetic activity is influencing cloud cover as this data seems to indicate then we expect some variable time delay between the geomagnetic variation and the resulting cloud variation. The data of Figure 5 indicate a small delay, about 0 to 3 days. There are times when cloud cover appears to be leading geomagnetic activity. However, synoptic cloud variation must have significant components in the pass band that will, occasionally, and especially near solstice, be dominant and mask the connection between geomagnetic activity and cloud.

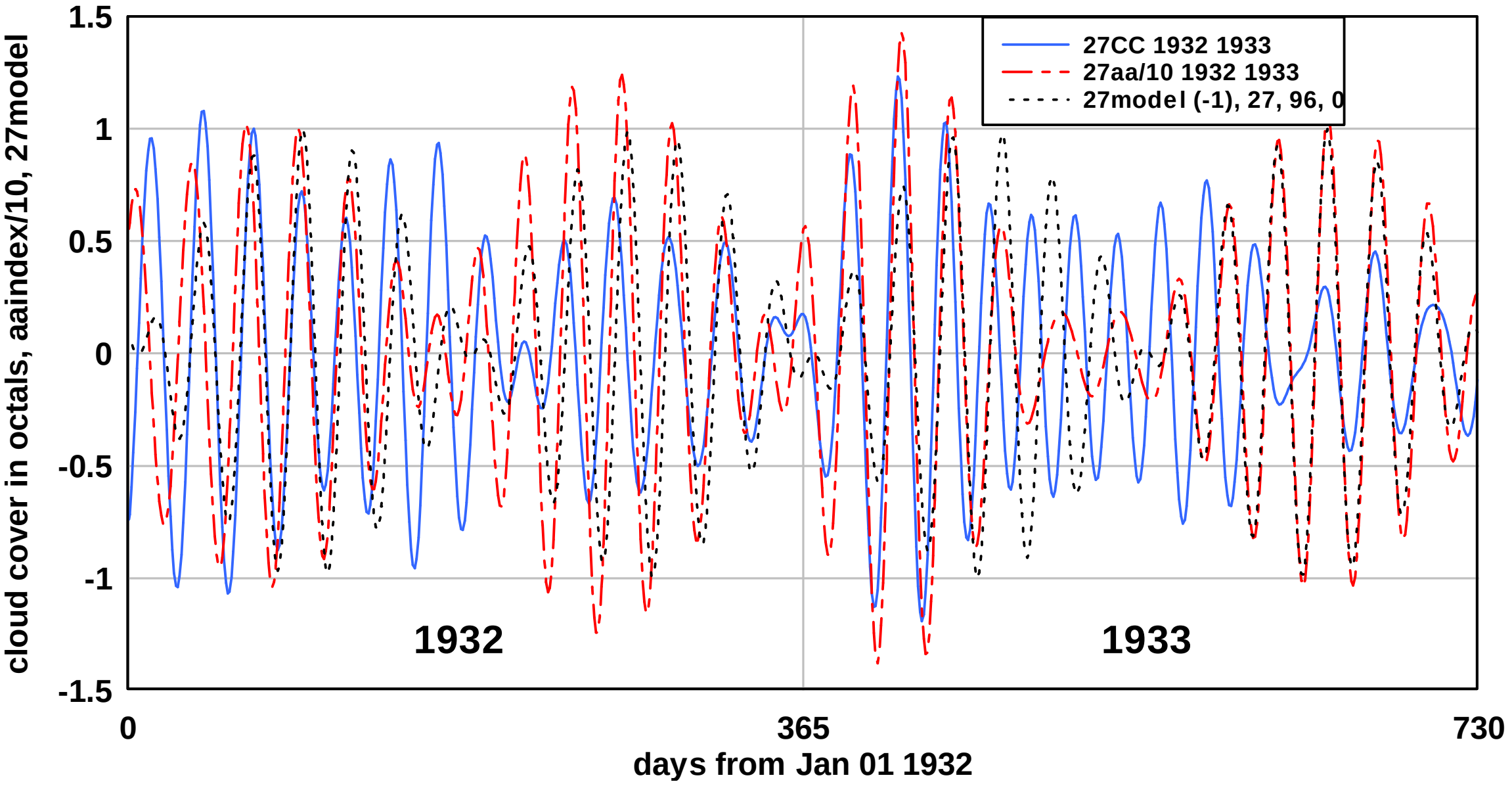

**Figure 5.** Compares the model variation with the variation of the ~27 day components of aa index and cloud cover at Hobart during 1932 1933.

The frequency spectra of the variations during 1932 – 1933 are shown in Figure 6. It should be noted that the spectral components of the band passed data in the frequency band 0.031 days$^{-1}$ and 0.045 days$^{-1}$ are the same as the frequency components obtained by frequency analysis of the raw (non-band pass) data. Also note that the point to point resolution provided by the 730 day record is 0.0014 days$^{-1}$. Despite the limited resolution it is clear that the spectrum of the aa index variation during 1932 1933 is close to an exact replica of the model spectrum. Notable is the fact that the central peak at 27 days has been entirely split into two sidebands at periods of 29.1 and 25.1 days, (indicated by the heavy vertical lines in Figure 6). Cliver et al (1996) obtained frequency spectra of the aa index over entire solar cycles and observed the splitting of the ~27 day component of aa index into two peaks at 25 and 29 days more strongly in even cycles than in odd cycles. Cliver et al (1996) attributed the splitting to the same amplitude modulation used by Coleman and Smith (1966) to interpret peaks in the Ci and Kp geomagnetic indices. As mentioned earlier the function used by Coleman and Smith (1966) is essentially the same as equations (2) or (3) derived here.

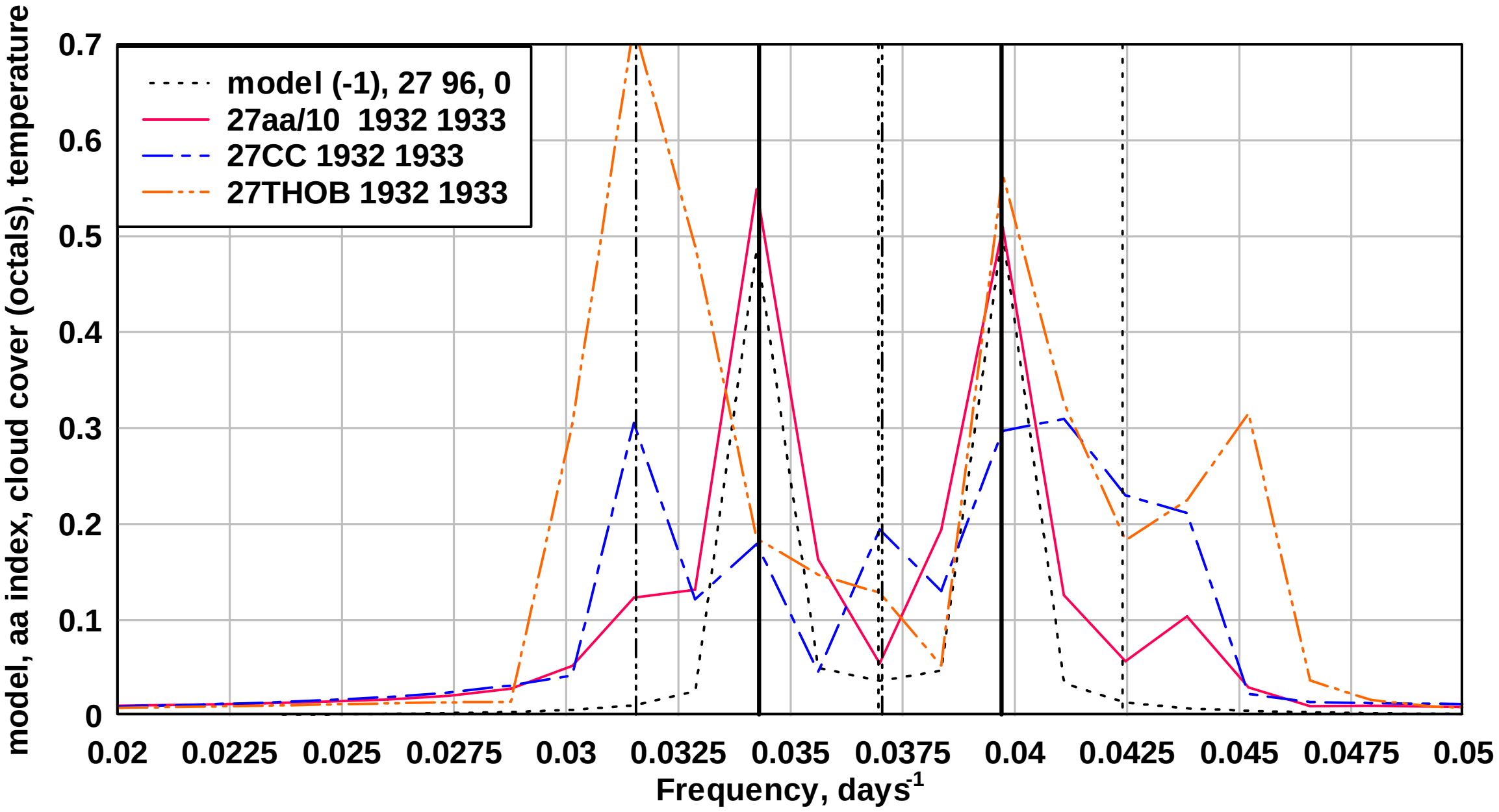

**Figure 6.** Compares the spectrum of the model variation with the spectrum of the ~27 component of aa index and the spectra of ~27 day components of cloud cover and maximum temperature at Hobart during 1932 1933. Reference lines are at 32, 29, 27(2), 25 and 23 day periods.

The spectrum of the cloud cover variation broadly follows the pattern of two peaks at 25 and 29 days. However, despite the limited resolution, it is clear that the peaks at 25 and 29 days are each further split into two sidebands shifted from the original side bands by +/- $f_A$ = +/- 0.00274 days$^{-1}$. Evidently this is due to a seasonal amplitude modulation of cloud variability. Thus six frequency components of the cloud cover variation can be expected at $f_S - 2f_A$, $f_S - f_A$, $f_S$, $f_S$, $f_S + f_A$ and $f_S + 2f_A$ or, approximately, at 32, 29, 27, 27, 25 and 23 days as indicated by the six vertical reference lines in Figure 6. The frequency spectrum of the daily maximum temperature variation at Hobart during 1932 1933, to be discussed in the next section, is also shown in Figure 6. The spectra of cloud cover variation during the other annual intervals studied are broadly similar to the spectra for cloud cover in the 1932 -1933 interval.

## 6. Comparison of the daily average aa index, cloud cover and surface temperature.

In the previous section we established that the ~27 day variation in aa index is, in the selected annual intervals, strongly correlated with the 27 day component of the model variation of geomagnetic activity developed in section 3. Therefore in this section we drop the model from the comparison and compare only the aa index, cloud cover and surface temperature. We have also previously established that the variations of the ~27 day components of aa index and cloud cover are, for most of the time, in phase. However, there are significant intervals when aa index and cloud are varying out-of-phase e.g. around July August in 1933 in Figure 5 and in June July in 1993 (supplement Figure 5B). This occasional anti-phase variation of aa index and cloud cover appears to occur mainly in winter. An explanation is not known. However, it may simply be that the semi-annual effect reduces the 27 day variation associated with geomagnetic activity to near zero at these times and synoptic weather variations within the ~27 day pass band become dominant. Cloud cover has a strong effect on surface temperature as cloud cover alters the balance between in-coming short wave radiation and out-going long wave radiation. The simple expectation is that an increase in cloud cover will usually decrease surface temperature in summer but may increase surface temperature in winter, i.e. there may be a $180^{o}$ phase shift in the response of surface temperature to cloud cover between summer and winter.

Figure 7 compares the ~27 day components of daily aa index, Hobart cloud cover and Hobart maximum temperature for 1932 1933. Hobart daily maximum temperature (THOB) 1918 to 2013 is available at http://www.bom.gov.au/climate/change/hqsites/data/temp/maxT.094029.daily.txt
Also shown is the 90 day running average of the daily maximum temperature anomaly at Hobart for 1932 1933. Readers will recall that a symmetrical variation in this anomaly and the crossing of zero by the anomaly near day 365 were used as criteria to select annual intervals for study. Clearly all the variables in Figure 7 show a semi-annual variation in amplitude with maxima near the equinoxes in 1932 and 1933. Of interest, however, is the variation in surface temperature. We notice that the variation in the ~27 day component of surface temperature is almost exactly in anti-phase with the ~ 27 day components of cloud cover and aa index at all times apart from the months July and August in 1933. This supports the simple expectation that an increase in daily cloud cover leads to a decrease in maximum surface temperature most of the time except, occasionally, in winter when an increase in cloud cover may lead to an increase in daily maximum temperature.

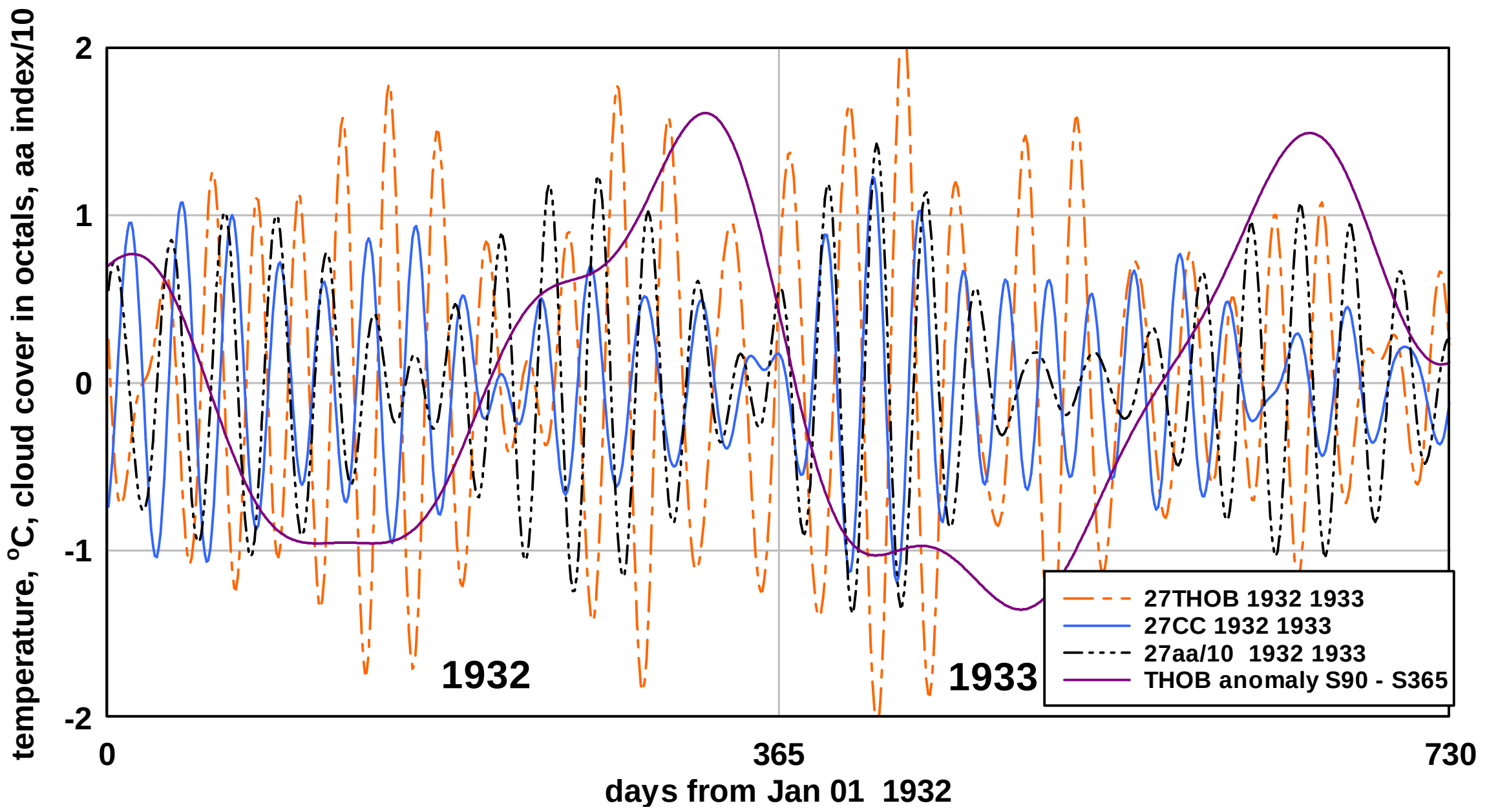

**Figure 7.** Compares the ~27 day components of aa index, cloud cover at Hobart and daily maximum temperature at Hobart during 1932 1933. Also shown is the 90 day running average of the maximum temperature anomaly at Hobart during 1932 1933.

Returning now to the variation in the 90 day average temperature anomaly also shown in Figure 7. This is an annual variation of order 3°C peak to peak remaining when the seasonal average temperature variation (in Hobart about 10°C peak to peak between summer and winter) is removed. The inference in this paper is that this annual variation in surface temperature anomaly is connected with the annual component in the model of geomagnetic activity illustrated in Figure 2A. We infer from the model that a connection between geomagnetic activity, cloud cover and surface temperature leads to a decrease in the average temperature anomaly around March equinox and an increase in the average temperature anomaly around September equinox. We note that the annual component in the model of geomagnetic activity results from including solar wind velocity (V) in the product $BV^2$ that we relate to geomagnetic activity. So it appears that the annual variation is due to the influence of solar wind velocity variation rather than magnetic field variation. However, further discussion of this aspect is outside the scope of the present paper.

## 7. Comparison of aa index and surface temperature at other locations.

The response of cloud and temperature to the semi-annual ~ 27 day period variation of geomagnetic activity as illustrated above is expected to be similar at other mid-latitude locations. However, recent reports have indicated the possibility of anti-phase variation between northern and southern hemisphere troposphere variables in response to geomagnetic activity, Burns et al (2008), Lam et al (2013), so in this section we study the response at northern hemisphere mid-latitude locations to compare with the response at Hobart. Daily maximum temperature data is fairly ubiquitous and accessible. However, accessible cloud data is much less common and it is necessary to make do with just temperature data in this section. We provide the ~27 day variations in daily average aa index and in daily maximum temperature at two northern hemisphere locations: Central

England, in Figure 8, and Boulder, Colorado (Figure 8B supplement). Daily maximum temperatures at these two locations are available at http://www.metoffice.gov.uk/hadobs/hadcet/cetmaxdly1878on_urbadj4.dat and http://www.esrl.noaa.gov/psd/boulder/data/boulderdaily.complete. We established in Section 5 that cloud cover at the southern hemisphere mid-latitude location Hobart varied mainly in-phase with the daily average geomagnetic aa index and, in Section 6, we established that, at Hobart, the ~27 day component of surface temperature varied, predominantly, out of phase with both the ~27 day aa index variation and the ~27 day cloud cover variation. Therefore at the climatically similar, but northern hemisphere, location of Central England we expect the ~27 day component of surface temperature to also vary predominantly in anti-phase to the ~27 day component of aa index unless the response in the two hemispheres differs significantly. The ~27 day variations of aa index and daily maximum temperature at Central England (27TCET) during 1932 1933, Figure 8, do vary in anti-phase apart from the months, January and June 1932 and November 1933 when the aa index and temperature variations vary in-phase. This suggests that the phase relationships of aa index, cloud cover and surface temperature at Central England are the same as they are at Hobart. However, there is an uncertainty due to the fact that surface temperature can vary both in-phase and out of phase with cloud cover (and therefore with aa index). This uncertainty cannot be resolved without access to northern hemisphere cloud cover records.

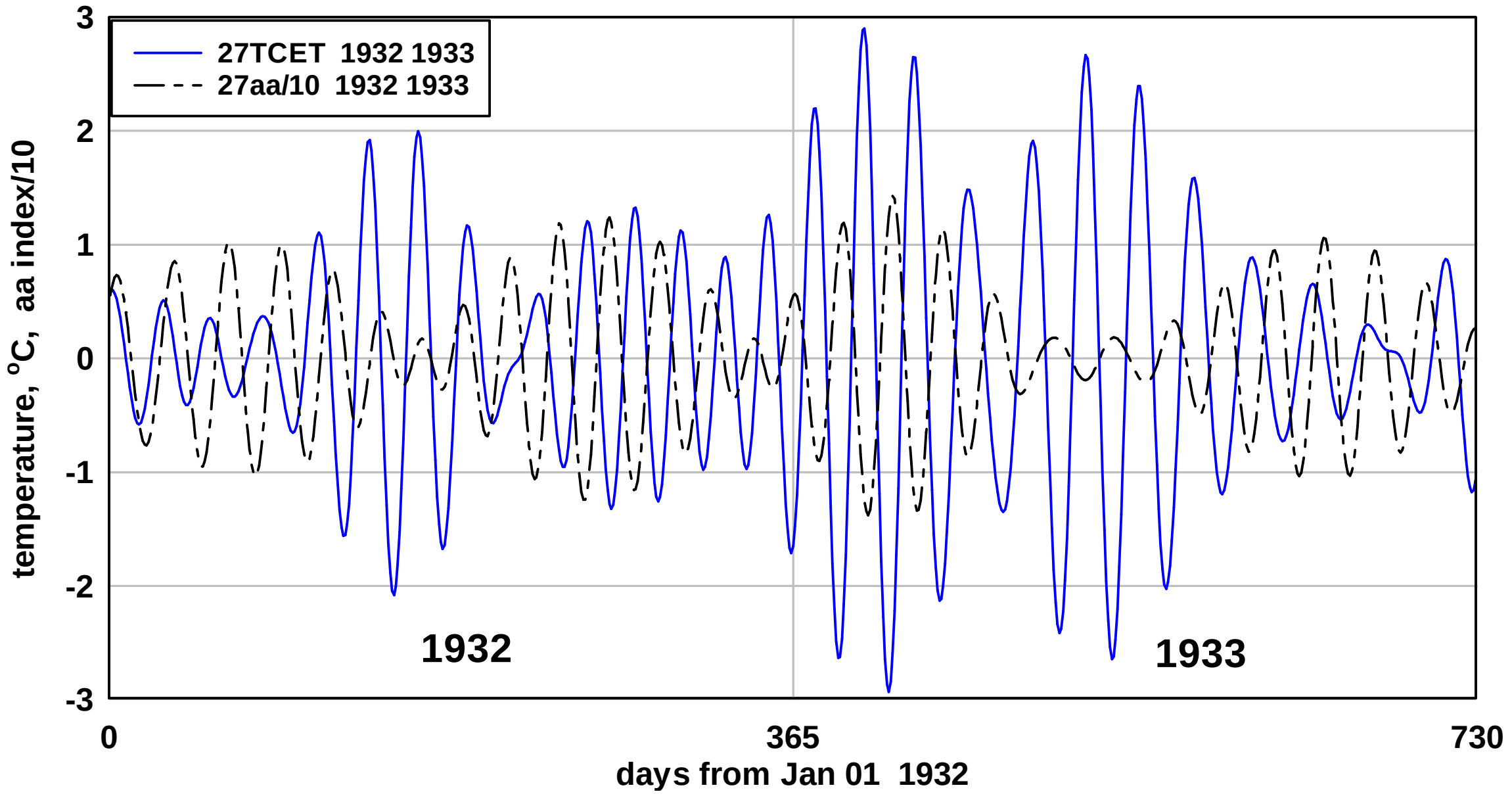

**Figure 8.** Compares the ~27 day components of daily average aa index and daily maximum temperature at Central England during 1932 1933.

Figure 9 compares the frequency spectrums of the ~27 day components of aa index, Central England surface temperature and Boulder, Colorado surface temperature (27TBOU) during 1932 1933. It was established previously, Figure 6, that, during 1932 1933, the spectrum of aa index closely replicates the two sideband peaks associated with the model. The spectra for 27TCET and 27TBOU broadly follow this two peak pattern. However, as discussed in section 5 and indicated in Figure 6, it is expected that the two sidebands in the temperature spectrum will be further split as a result of seasonal modulation of both the cloud variability and the temperature variability. Thus components at $f_S$ +/- $nf_A$ are expected where n is the number of modulations including the modulation due to the semi-annual effect. With n = 3 the frequencies of the two most widely split components, one at 0.029 days$^{-1}$ (34 days) and the other at 0.045 days$^{-1}$ (22 days) are, respectively, beyond and on the limits of the band pass filter used in this study, (0.031 to 0.045 days$^{-1}$). However the spectra in Figure 9 support the concept of two sidebands at 25 and 29 days being further split by seasonal amplitude or phase modulation of cloud cover and surface temperature.

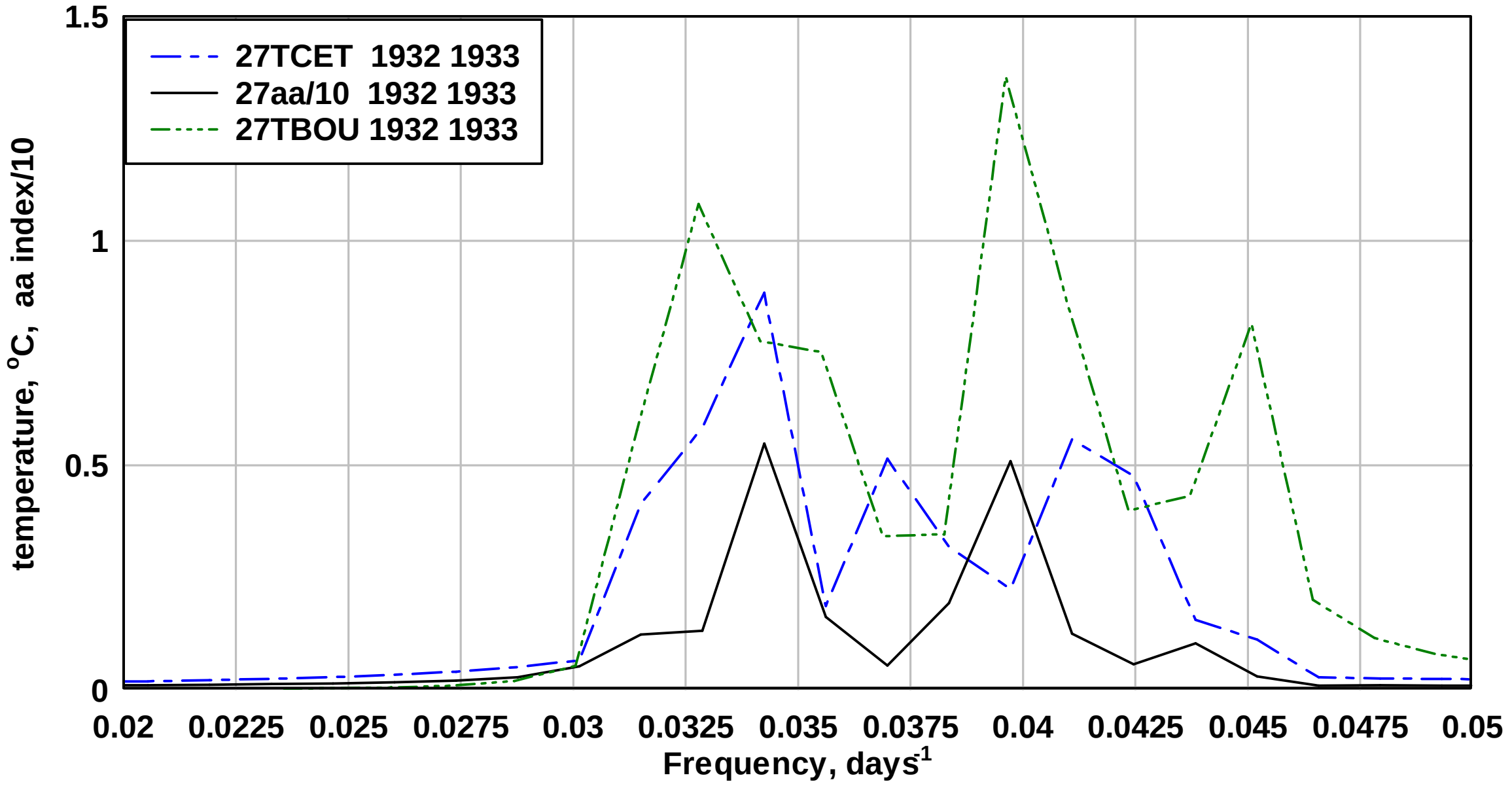

**Figure 9.** Compares the frequency spectrum of the ~27 day component of aa index with the spectra of the ~27 components of daily maximum temperature at Central England and Boulder, Colorado during 1932 1933.

## 6. Discussion and conclusions.

In this paper we inferred a chain of connection between (1), the variation of the solar magnetic field and the solar wind velocity at Earth as determined by the Sun's rotation and dipole tilt and the heliographic location of Earth; (2), geomagnetic activity; (3), cloud cover; and (4), surface temperature. For the first step in the chain, (1) - (2), the model of geomagnetic activity derived from the time varying geometry of the Sun and Earth was successful in predicting the time variation of the geomagnetic aa index in selected years. For example in the years 1932 1933 the correlation between the 27 day component of the model and the ~27 day component of the aa index was 0.77. The model accurately predicts the strong semi-annual effect in the ~27 day component of geomagnetic activity, with strongest effect at equinox and with weakest effect at solstice. Thus the expectation is that near equinox the correlation would be higher than the average correlation obtained over the entire year. This has not been quantified but it is evident, for example from Figures 4 and 5, that this expectation is supported. At times near solstice it is expected that, aside from the approach to zero variation, the variation in geomagnetic activity at this time would be unrelated to solar rotation due to the effect of synoptic weather. Therefore, at solstice it is expected that the the 27 day components of aa index, cloud and temperature would not exhibit a steady ~27 day variation. The observed annual correlations between the model and the aa index should therefore be viewed as arising from only half of the interval of the correlation, the half of the interval around the equinoxes.

The second step in the chain, (2) – (3); the connection of geomagnetic activity to cloud cover should be viewed in the same context: that significant correlation between the two variables is expected only during times around equinox. Again, this expectation has not been quantified but is supported by the observations in Figure 4 and Figure 5. We conclude that there is a significant semi-annual effect of geomagnetic activity on cloud cover. It is difficult to work out how significant this contribution is. For example, Figure 10 shows that the ~27 day component of cloud cover amounts to about 25% of the overall variation of cloud cover during the interval 1932 1933. However, from the above discussion, significant geomagnetic influence is expected during only half of this two year interval, during the twelve months around the equinoxes, and we expect much less geomagnetic influence around solstices. So part of the challenge is how to partition and quantify the influence. One possibility is to compare correlations made in the months around equinox with correlations made in the months around solstice. However, this would only work for semi-annual events symmetrical about January 01 as were chosen for this study. It is evident from Figure 3 and Figure 3B and 3C in the supplement that many of the semi-annual events, particularly events during solar maximum are neither symmetrical nor symmetrical about January 01. Nevertheless, the expectation of zero correlation around solstice helps to explain the observation of, apparently, anti-phase variation of ~27 day aa index and cloud cover that was noted to occur for a few months near solstice in the variations discussed in Section 5 and illustrated in Figures 4 and 5.

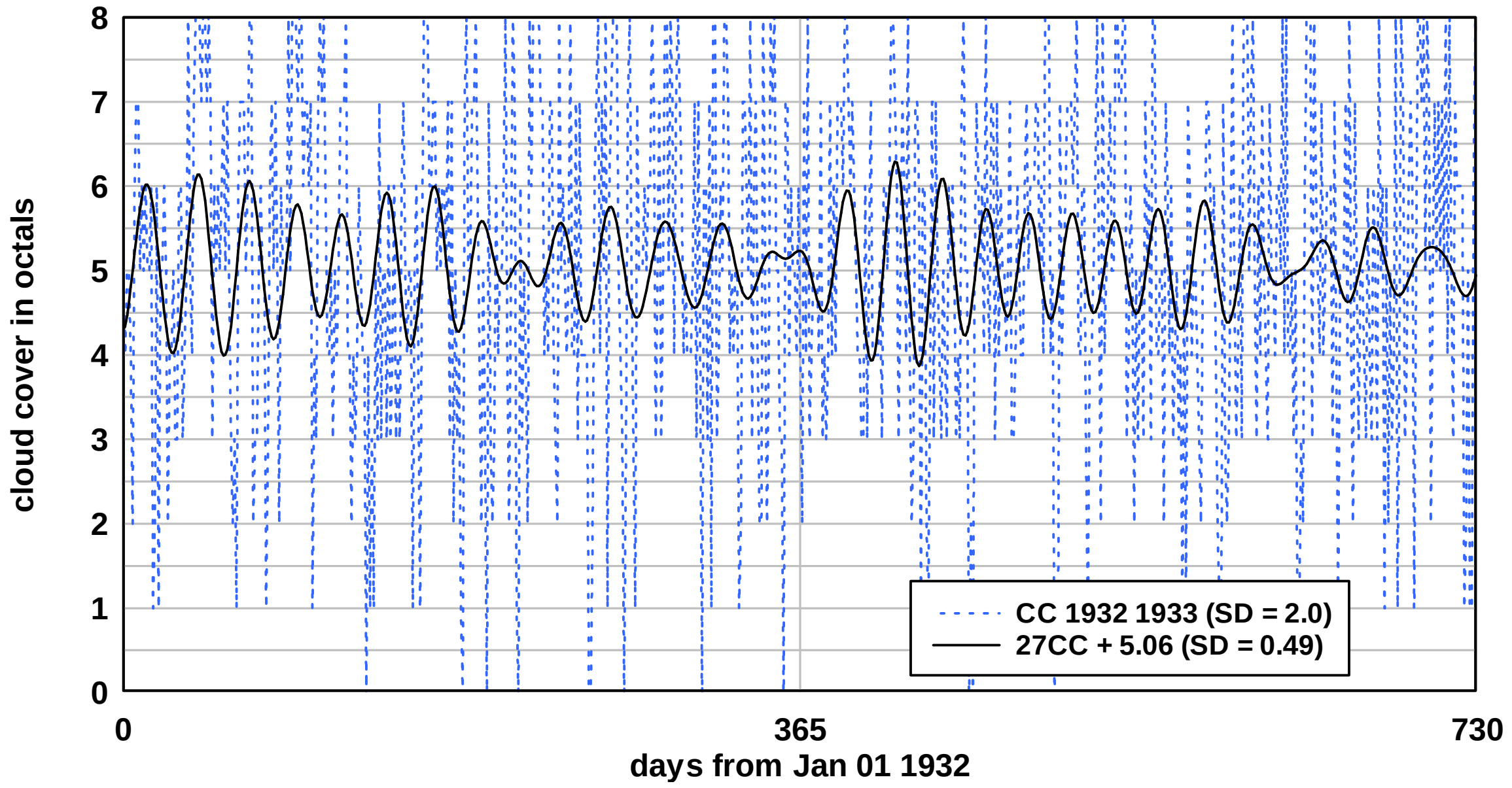

**Figure 10.** Compares the overall cloud cover variation with the ~27 day component of the cloud cover variation for Hobart during 1932 1933. The standard deviation of cloud cover is 2.0 octals and the standard deviation of the ~27 day component of cloud cover is 0.49 octals.

The third step in the chain, (3) – (4), the connection between cloud cover and temperature is less problematic. It is well known that there is usually an inverse relationship between daily average cloud cover and daily maximum surface temperature. Section 5 showed that the ~27 day variation in cloud cover varies, for most of the time, in-phase with the ~27 day variation of the aa index. Therefore, the expectation is that the ~27 day variation in surface temperature should vary in anti-phase the ~27 day component of aa index. This was clearly supported by the observations in Figure 7 and Figure 8. We conclude therefore that, at times around equinox, in the selected years studied, the surface temperature is influenced by variations in geomagnetic activity. Again it is difficult to work out how significant the contribution is. During 1932 1933, after removing the seasonal, summer to winter variation, about 30% of the residual variation is accounted for by the ~27 day variation, Figure 11 (supplement). However, as some of this ~27 day variation is contributed from synoptic weather, certainly near solstice, there is again the difficulty of estimating how much of the ~27 day variation is due to geomagnetic activity.

This study could be extended in various ways. The model of magnetic field variation could be developed to include the effect of equatorial solar dipoles. The influence of geomagnetic activity on cloud cover could be examined at other locations by accessing cloud cover records or extending the study to sunshine duration and solar exposure records. As it appears that a significant influence of ~27 day geomagnetic activity on cloud and temperature predictably occurs during years around solar minimum and at times around equinox it would be interesting to examine if this could assist in long term weather prediction. The solar magnetic field and the solar wind velocity contain significant components at harmonics of the 27 day solar rotation period e.g $f_S/2$, or 13.5 day period. It would be interesting to examine if cloud cover and surface temperature contained components at $f_S/2$ +/- $nf_A$.

## Supplementary material.

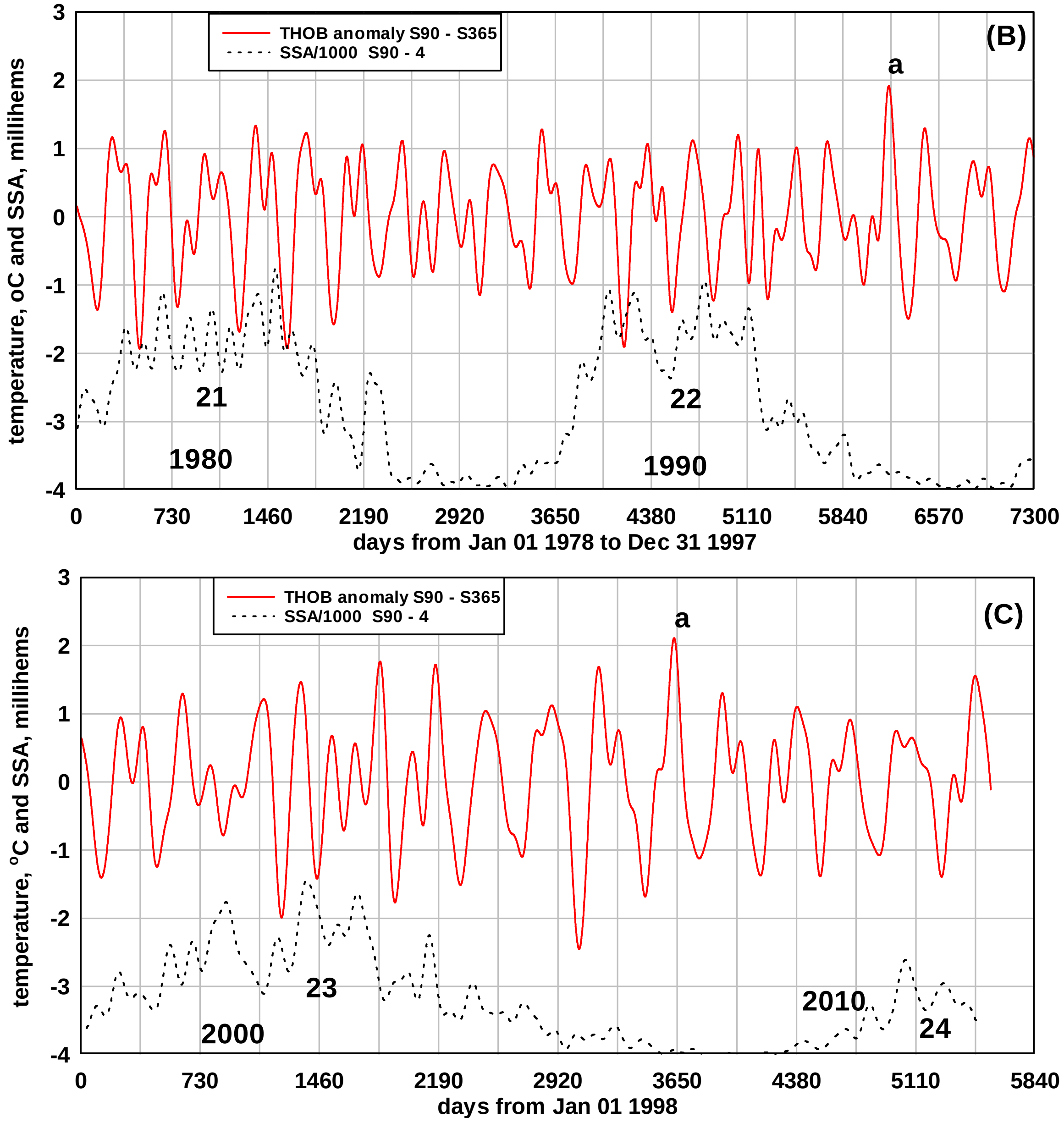

**Figure 3B and 3C.** Shows the 90 day running average of the daily maximum temperature anomaly (THOB) at Hobart for the years 1978 – 1987 (B) and years 1998 – 2013 (C).

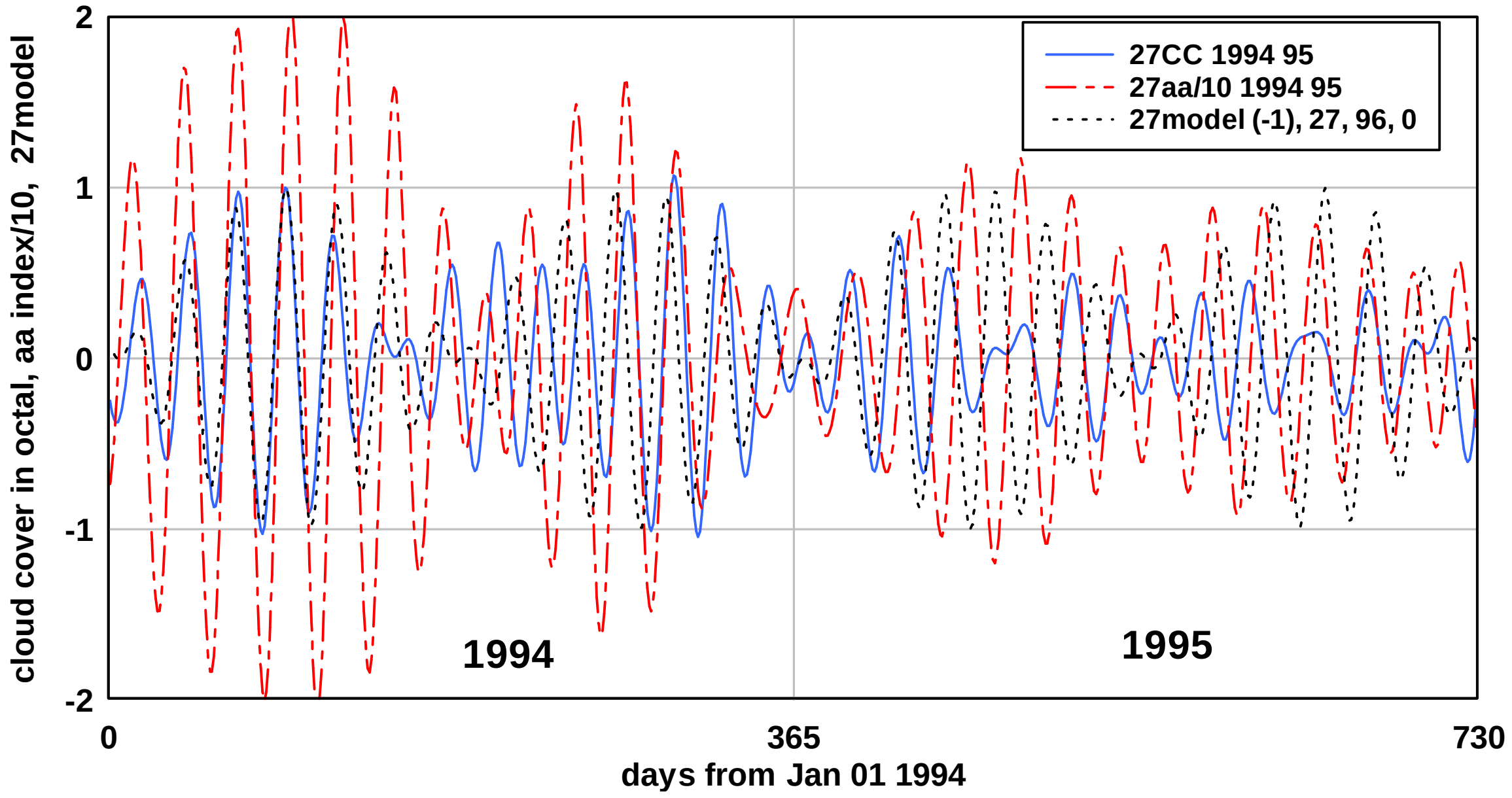

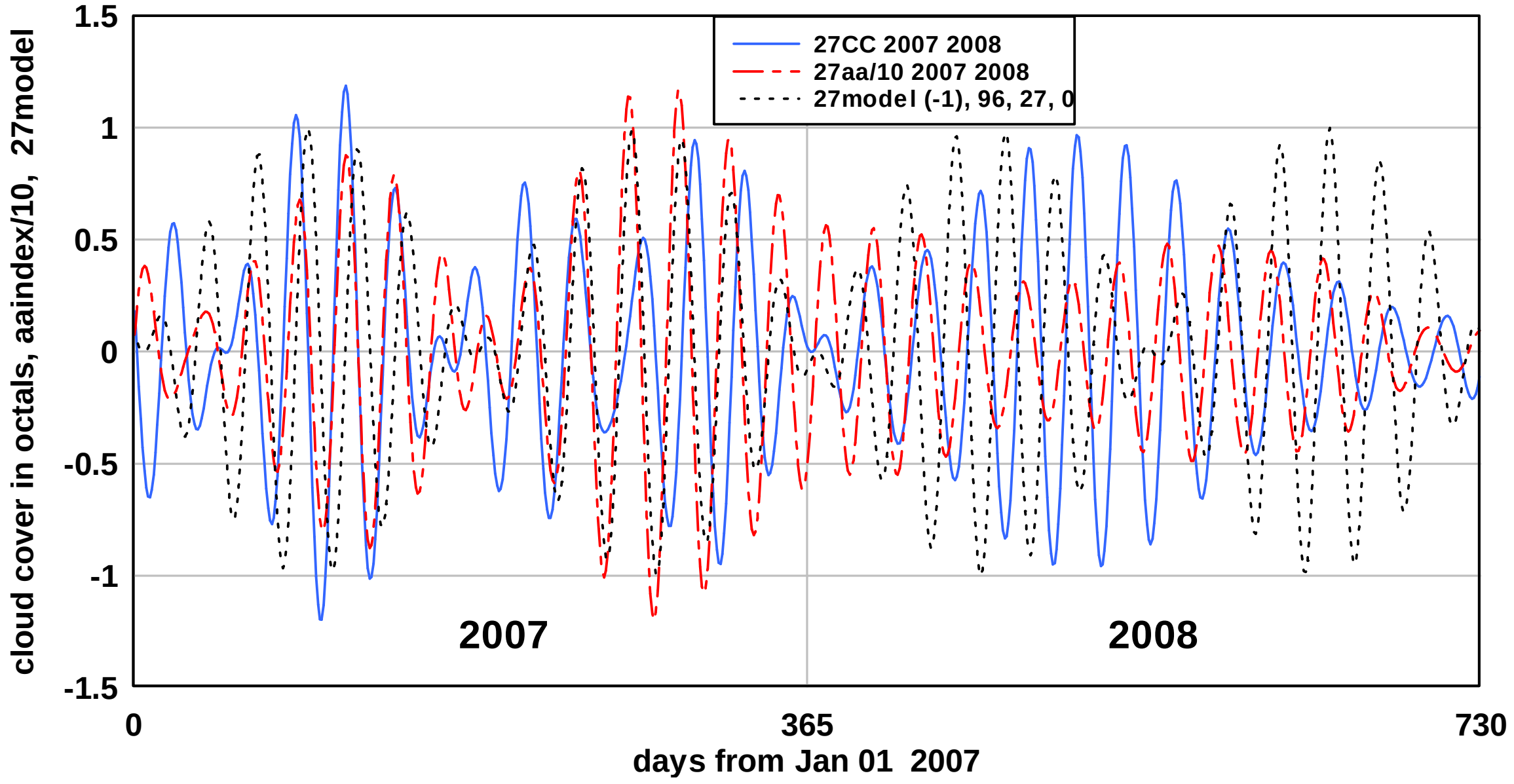

**Figure 5B and 5C.** Compares the model variation with the variation of the ~27 day components of aa index and cloud cover at Hobart during 1994-1995 and 2007-2008.

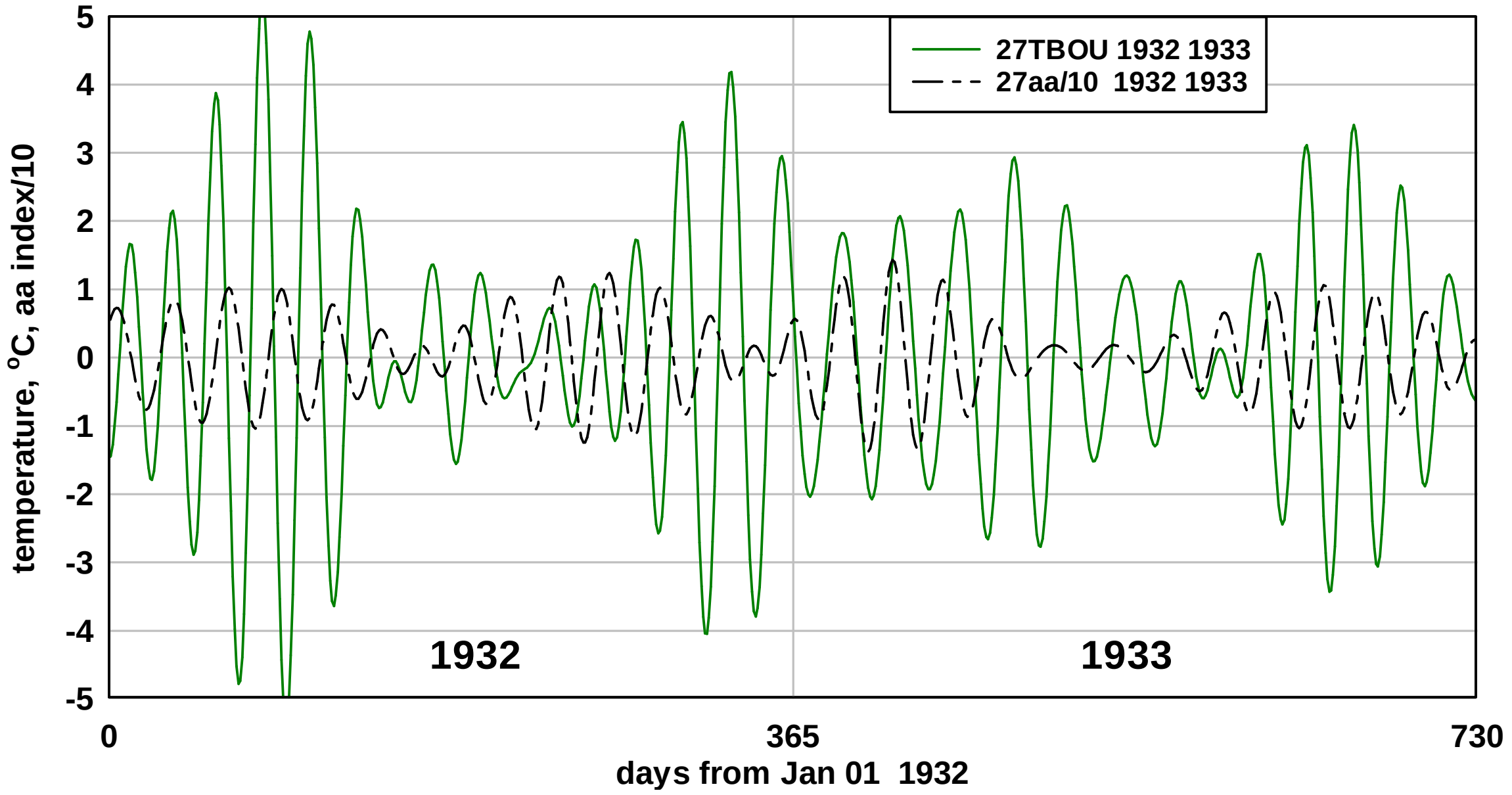

**Figure 8B.** Compares the ~27 day components of daily average aa index and daily maximum temperature at Boulder, Colorado during 1932 1933.

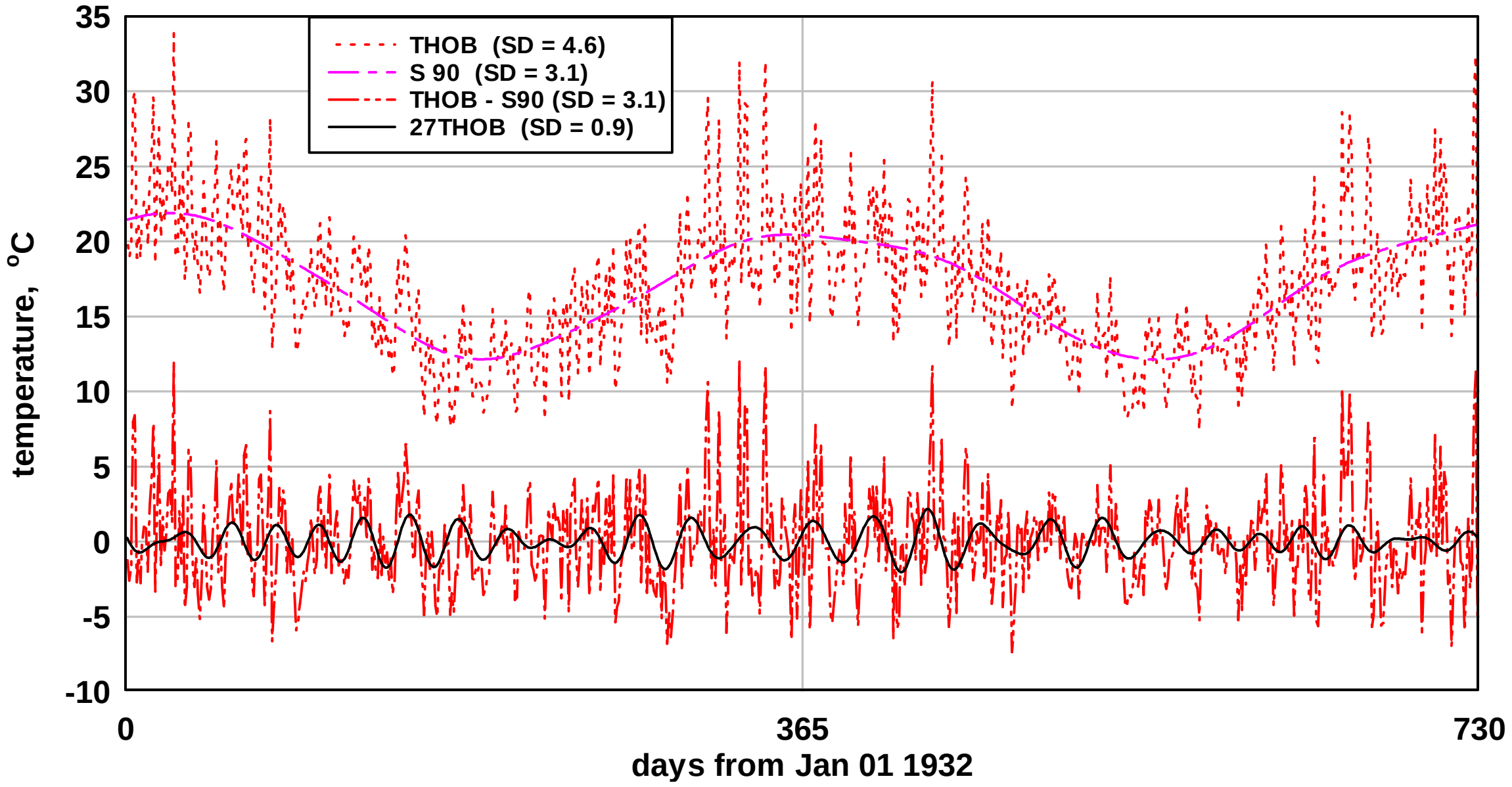

**Figure 11.** Compares the residual daily maximum temperature variation after the seasonal variation has been removed (THOB – S90) with the ~27 day component of the daily maximum temperature (27THOB). The standard deviations of each variable are indicated in the legend.